\documentstyle[twoside,fleqn,espcrc2,epsfig,graphics]{article}

\def\benum{\begin{enumerate}}
\def\eenum{\end{enumerate}}

\def\ben{\begin{equation}}
\def\ba{\begin{array}}
\def\bea{\begin{eqnarray}}
\def\bec{\begin{center}}

\def\een{\end{equation}}
\def\eea{\end{eqnarray}}
\def\enc{\end{center}}
\def\ea{\end{array}}
\def\btab{\begin{table}}
\def\btabu{\begin{tabular}}
\def\etab{\end{table}}
\def\etabu{\end{tabular}}
\def\bit{\begin{itemize}}
\def\eit{\end{itemize}}
\def\bef{\begin{figure}[tb]}
\def\befh{\begin{figure}[!h!]}
\def\enf{\end{figure}}

\def\Sc{\scriptstyle}

\def\la{\langle}
\def\ra{\rangle}

\def\a{\alpha}

\def\gb{\beta}

\def\V{{\cal V}}

\def\s{\sigma}

\def\b1{{\bf 1}}

\def\bS{{\mbox{\boldmath $S$}}}

\def\bx{{\mbox{\boldmath $x$}}}
\def\sbx{{\mbox{\boldmath $\Sc x$}}}

\def\sbmu{{\mbox{\boldmath $\Sc\mu$}}}
\def\sbnu{{\mbox{\boldmath $\Sc\nu$}}}
\def\bmu{{\mbox{\boldmath $\mu$}}}
\def\bnu{{\mbox{\boldmath $\nu$}}}

\def\nn{\nonumber}
\def\bb{\left(}
\def\eb{\right)}

\def\bem{\begin{minipage}}
\def\enm{\end{minipage}}

\newcommand{\AmS}{{\protect\the\textfont2
  A\kern-.1667em\lower.5ex\hbox{M}\kern-.125emS}}

\title{
Monte-carlo renormalization group study of gauged $RP^2$ 
 spin models in two dimensions
\thanks{Presented by R.R. Horgan. This work is supported by NATO collaborative 
research grant no. CRG950234 and in part by the Leverhulme Trust.}
}

\author{
S.M. Catterall\address{Department of Physics, Syracuse University,
                       Syracuse, NY 13244-1130, USA}
M. Hasenbusch\address{Fachbereich Physik, Humbolt Universit\"at zu Berlin,
                       Invalidenstr. 110, 10099 Berlin, Germany}
R.R. Horgan\address{DAMTP, University of Cambridge, Silver Street, 
                     Cambridge, England CB4 4SL} 
R. Renken\address{Department of Physics, University of Central Florida, 
                   Orlando, FL32816, USA} 
}

\begin{document}
\begin{abstract}
The 2D $RP^2$ gauge model is studied using the Monte-Carlo Renormalization
Group (MCRG). We confirm the first-order transition reported in
\cite{SSD} ending in a critical point associated with
vorticity. We find evidence for a new renormalized trajectory (RT) which is 
responsible for a cross-over from the vortex dominated regime to the $O(3)$ regime 
as the coupling is reduced.  Near to the cross-over region a good signal 
for scaling will be observed in $RP^2$ but this is illusory and is due to the 
proximity of the RT.  We suggest that this is the origin of the `pseudo'-scaling 
observed in \cite{haho}. We find that the continuum limit of $RP^2$ is controlled 
by the $O(3)$ fixed point.
\end{abstract}

\maketitle

\section{Introduction}
\enlargethispage{-8truemm}
In \cite{haho} the mass gap measurement in the $SO(4)$ matrix model
was compared with the Bethe-Ansatz prediction and found to disagree
by a factor of four although for the covering group model there
was good agreement. It was concluded that the signal
for scaling in the $SO(4)$ simulation was only apparent and that a true
continuum limit had not been achieved. It was conjectured that 
the deception was due to vortices, present in $SO(4)$ but
absent in the covering group. Here we investigate whether vortices can cause
a bogus signal for scaling in the 2D $RP^2$ gauge model which allows an interpolation 
between the pure $RP^2$ and the $O(3)$ spin models and which also contains vortices.

We find evidence for the $O(3)$ fixed point and for a scaling flow 
which indicates the presence of a new renormalized trajectory (RT).
We suggest that this RT will cause an apparent scaling signal of the kind reported
in \cite{haho} and that it is responsible for an observed cross-over effect.

It has also been 
conjectured \cite{Sokal} that in 2D the continuum limit in pure $RP^2$ 
is distinct from that in pure $O(3)$. Niedermayer et al. \cite{Nieder}
and Hasenbusch \cite{Has} have suggested that this conjecture is incorrect and
that the continuum limit in the $RP^2$ model is controlled
by the $O(3)$ fixed point. The work we report here supports this conclusion. 

\enlargethispage{-8truemm}
%\section{The model}
The action used is
\ben
S~=~-\gb\bb \sum_{\sbx,\sbmu}\,\bS_\sbx\cdot\bS_{\sbx+\sbmu}\,\s_{\sbx,\sbmu}~+~
    \mu\,\sum_P\,P(\s) \eb, \label{ACTION}
\een
where $\bS_\sbx$ is a unit length three-component vector at site $\bx$ and
$\s_{\sbx,\sbmu}$ is a gauge field on the link $\bx,\bmu$ taking values
in $[1,-1]$~. The plaquette of gauge fields is denoted by $P(\s)~$ and vortices
reside on plaquettes where $P(\s)=-1$. The pure $O(3)$ and $RP^2$ spin models
correspond to $\mu \rightarrow \infty$ and $\mu=0$ respectively.

%\section{The simulation}

A local update was used comprising a combination of heat-bath, microcanonical
and demon schemes. Lattice sizes ranged from $64^2$ to $512^2$ with typically 
$\sim 5\cdot 10^5$ configurations per run. The simulations were carried out on 
the HITACHI SR2201 computers in the Cambridge High Performance Computing Facility 
and in the Tokyo Computing Centre.

\section{The MCRG scheme}

To establish the topology of RG flows we study how the mean values of the spin-spin 
interaction, $A$, and of the plaquette, $P$, flow under blocking. For a given 
configuration these operators are given by
\bea
A~&=&~{1\over 2V}\sum_{\sbx,\sbmu}\,\bS_\sbx\cdot
      \bS_{\sbx+\sbmu}\,\s_{\sbx,\sbmu}~,\nn\\
P~&=&~{1\over V}\sum_P\,P(\s)~. \label{AP}
\eea
$\la A\ra \in [0,1]~, \la P\ra \in [-1,1]$ and the vorticity is $\V = (1-\la P\ra)/2~$. 

For each configuration $\{\bS,\s\}$ on a lattice of side $L$ we derive a
blocked configuration $\{\bS^B,\s^B\}$ on a lattice of side $L/2$~. The
blocking transformation for the spins is
\[
\bS^B_{\sbx_B}~=~\left(\bS_\sbx~+~\a\sum_\nu\,\bS_{\sbx+\sbnu}\s_{\sbx,\sbnu}\right)
                 /\left|\ldots\right|~.
\]
where $\bnu$ runs over nearest neighbour displacements from $\bx$, and
$\a$ was chosen to be 0.0625~. To block the links the gauge field
products were computed for the three shortest Wilson paths $W_0, W_+, W_-$
joining  the end points of the blocked link. The blocked gauge field
was assigned the majority sign of the $W_i$~.  

%\section{The measurement}
For given coupling constants $(\gb,\mu)$ and given lattice size $L^2$
each configuration was blocked by successive transformations until the
blocked lattice was $8^2$. The operators $A_L(\gb,\mu)$ and $P_L(\gb,\mu)$ 
were then measured and averaged over all configurations. This was done
for $L=64,128,256,512$ giving a flow segment in the $(\la A\ra,\la P\ra)$ 
plane with each point labelled by initial lattice size. The flow can be 
extended by tuning new couplings $(\gb^\prime,\mu^\prime)$ so that
\bea
A_{L^\prime}(\gb^\prime,\mu^\prime)~&=&~A_L(\gb,\mu)~,\nn\\
P_{L^\prime}(\gb^\prime,\mu^\prime)~&=&~P_L(\gb,\mu)~,\label{TUNE}
\eea
for some $L$ and $L^\prime$. The segment for fixed $(\gb^\prime,\mu^\prime)$ can then 
be computed. These flows result from a projection onto the $(\gb,\mu)$ plane
of true trajectories in a higher-dimensional coupling constant space. The assumption
is that only $A$ and $P$ correspond to relevant operators and that by blocking
the effect of irrelevant variables is minimized.

\section{Results and Conclusions}
There is a line of first-order transitions, first reported in \cite{SSD}, for $\mu \sim -0.3$ and
$\gb \ge 7.5$. We find that this line of transitions ends at the ``vorticity'' critical point. 
In figure \ref{f1} we plot $\la A\ra$ and $\la P\ra$ versus $\mu$ for $\gb = 8.0$. The order 
parameter is the vorticity. The linear combination $U = 2A + \gamma P$ with $\gamma \sim -0.29$ 
shows {\it no} discontinuity in this region. $U$ is closely related to the action, $S$ 
(eqn. \ref{ACTION}), and it is reasonable to interpret $U$ as the free energy. Also, $U$ is 
the correct operator to interpolate the vector state associated with the $O(3)$ critical
surface where the corresponding correlation length will diverge but which remains {\it finite}
at the vorticity critical point.

\bef
\bec
\epsfig{file=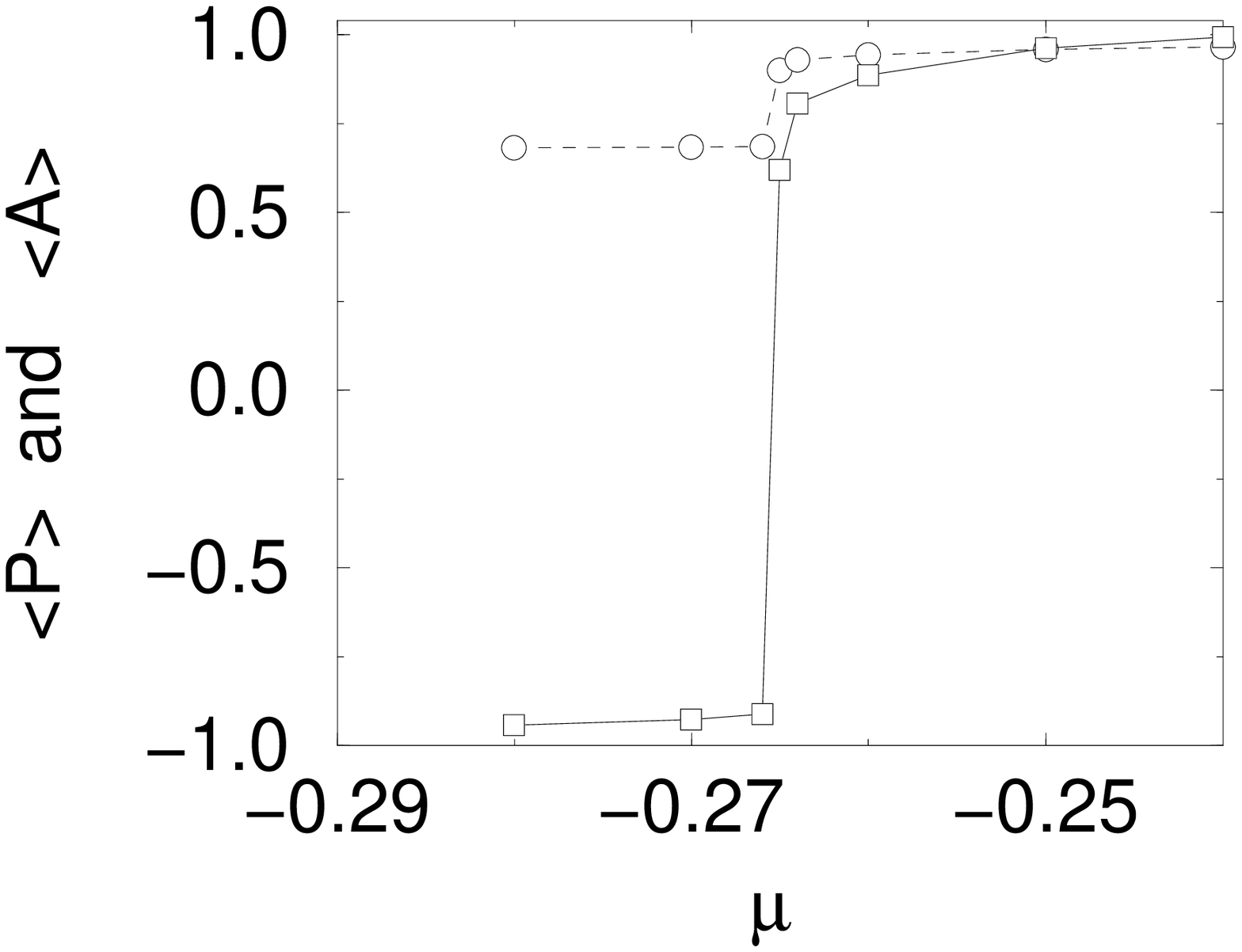,height=58mm}
\enc
\vskip -12truemm
\caption{\label{f1} $\la A\ra$ (dashed) and $\la P\ra$ (solid) versus $\mu$
for $\gb=8.0$ showing the first-order transition.}
\vskip -5truemm
\enf

There is a set of neighbouring flows on which the observables scale. The example
shown in figure \ref{f2} shows that there is a RT in the vicinity which must be associated 
with a new fixed point. To see that observables scale each flow segment of four points was 
successively overlaid using eqn. (\ref{TUNE}) with $L^\prime = L/2$. For the scaling 
flows the points of the overlaid flow segments coincide very well within errors.  
The flow is consistent with an exponent $\nu = 4.75(15)$ with critical couplings $(\gb^*,\mu^*)
\approx (7.5,-0.3)$. Of course, the new fixed point will not lie in the $(\gb,\mu)$ plane
and its relationship with the vorticity critical point is unclear. Open questions are: can the 
continuum limit of the new RT be taken at the vorticity critical point and does there exist 
a continuum limit with a non-zero vorticity density? It is intriguing that the position of the
vorticity critical point and the critical couplings deduced from the fit for $\nu$ are very 
similar. The nature of the new fixed points needs more investigation.

\bef
\bec
\epsfig{file=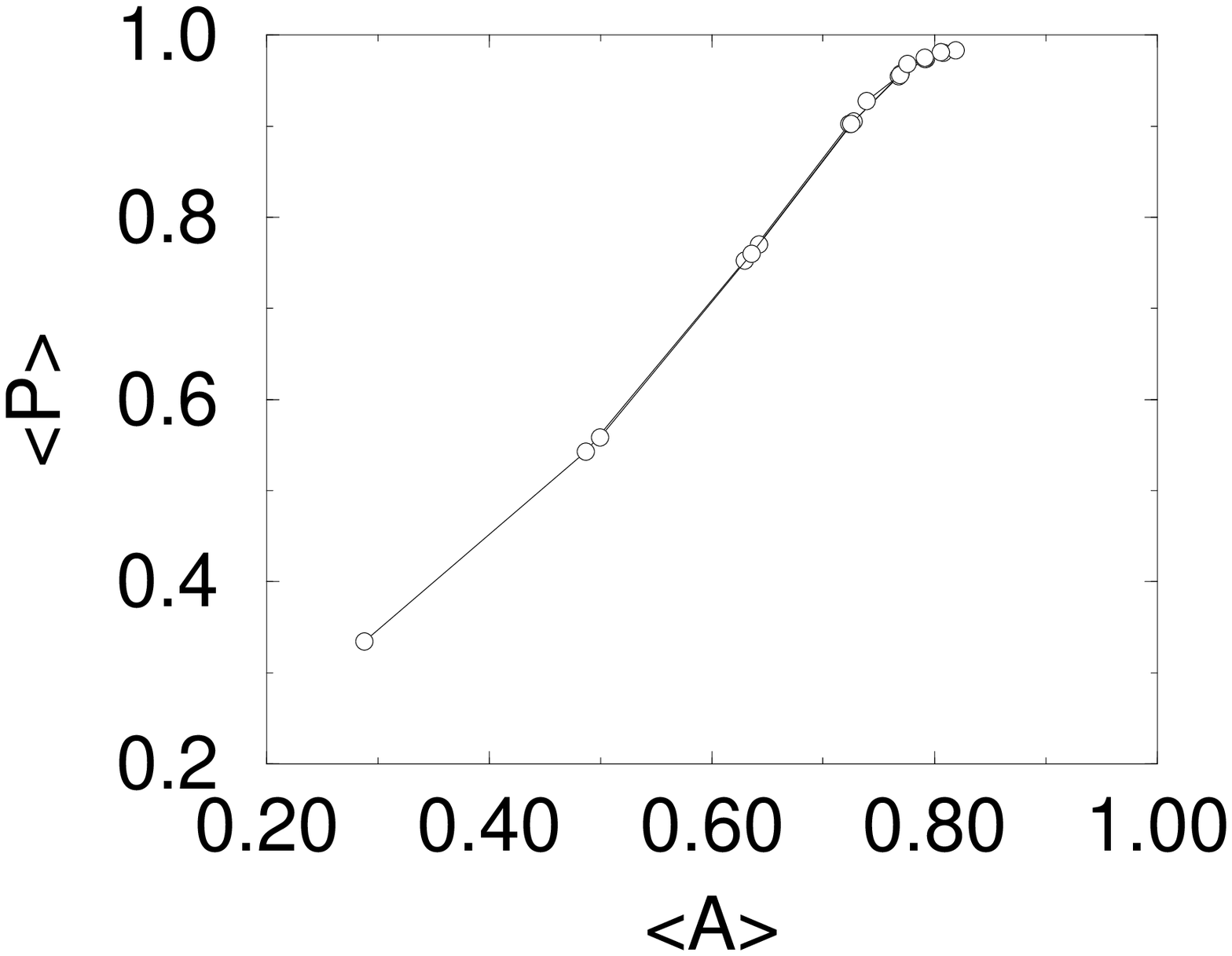,height=58mm}
\enc
\vskip -12truemm
\caption{\label{f2} Scaling flow. All points are from the $8^2$
blocking of, respectively, $L = 64,128,256,512~$.} 
\vskip -5truemm
\enf

Scaling on the $O(3)$ RT ($\mu=\infty$) was verified and agreement with the 3-loop perturbative 
$\gb-$function was found after effects due to finite $L$ were accounted for using 
perturbation theory.

The pure $RP^2$ model ($\mu=0$) does not intersect any critical surface except the one 
controlled by the $O(3)$ fixed point at $(\gb,\mu) = (\infty,\infty)$ ($\,(\la A\ra,\la P\ra) = (1,1)\,$). 
This confirms the conjectures of \cite{Nieder,Has} that $RP^2$ and $O(3)$ have the 
same continuum limit. In figure \ref{f3} flow segments for the $RP^2$ are 
shown for various $\gb$ in the range $3.9$ to $4.5$. There is a clear cross-over 
in the blocked vorticity, $\V$, as $\gb$ increases through this range. Moreover, the flows to
larger $\V$ renormalize close to the scaling flow, shown in figure \ref{f3} as
a dotted line. Hence, for $\gb \sim 4.0$ we should expect to see a good signal for scaling 
induced by the associated new RT. However, this scaling is not a signal for a continuum limit in $RP^2$ 
but is due to the proximity of the cross-over region to the new fixed point to which it is due. 
As $\gb$ is increased through the cross-over region scaling will be violated and 
eventually reappear in association with the true continuum limit controlled by the 
$O(3)$ fixed point. We believe that this is the origin of the bogus scaling observed in \cite{haho}.

\bef
\bec
\epsfig{file=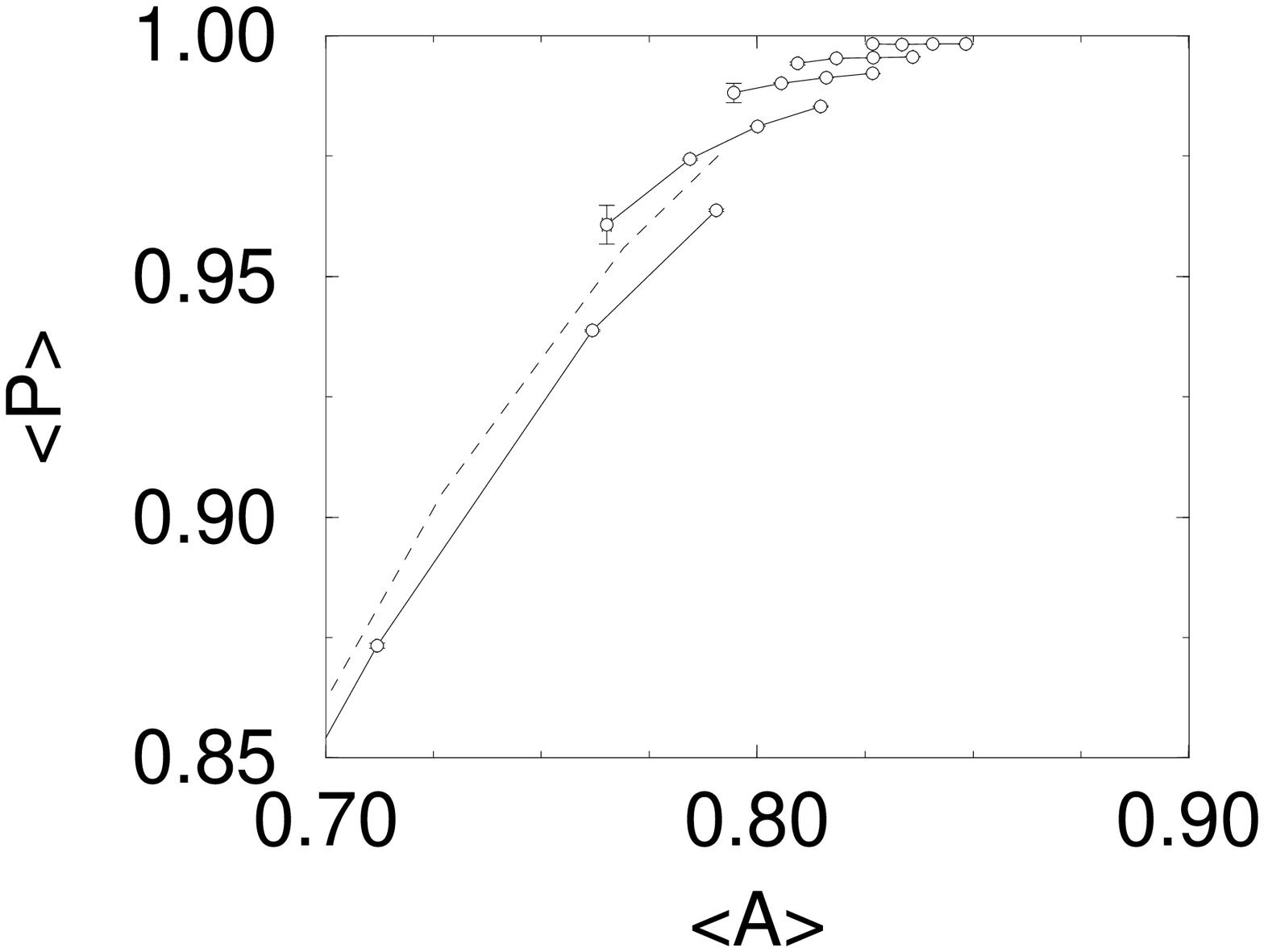,height=58mm}
\enc
\vskip -12truemm
\caption{\label{f3} The cross-over region in $RP^2$ ($\mu=0$) for 
$\gb = 3.9, 4.05, 4.17, 4.29, 4.5$.}
\vskip -5truemm
\enf

\befh
\bec
\epsfig{file=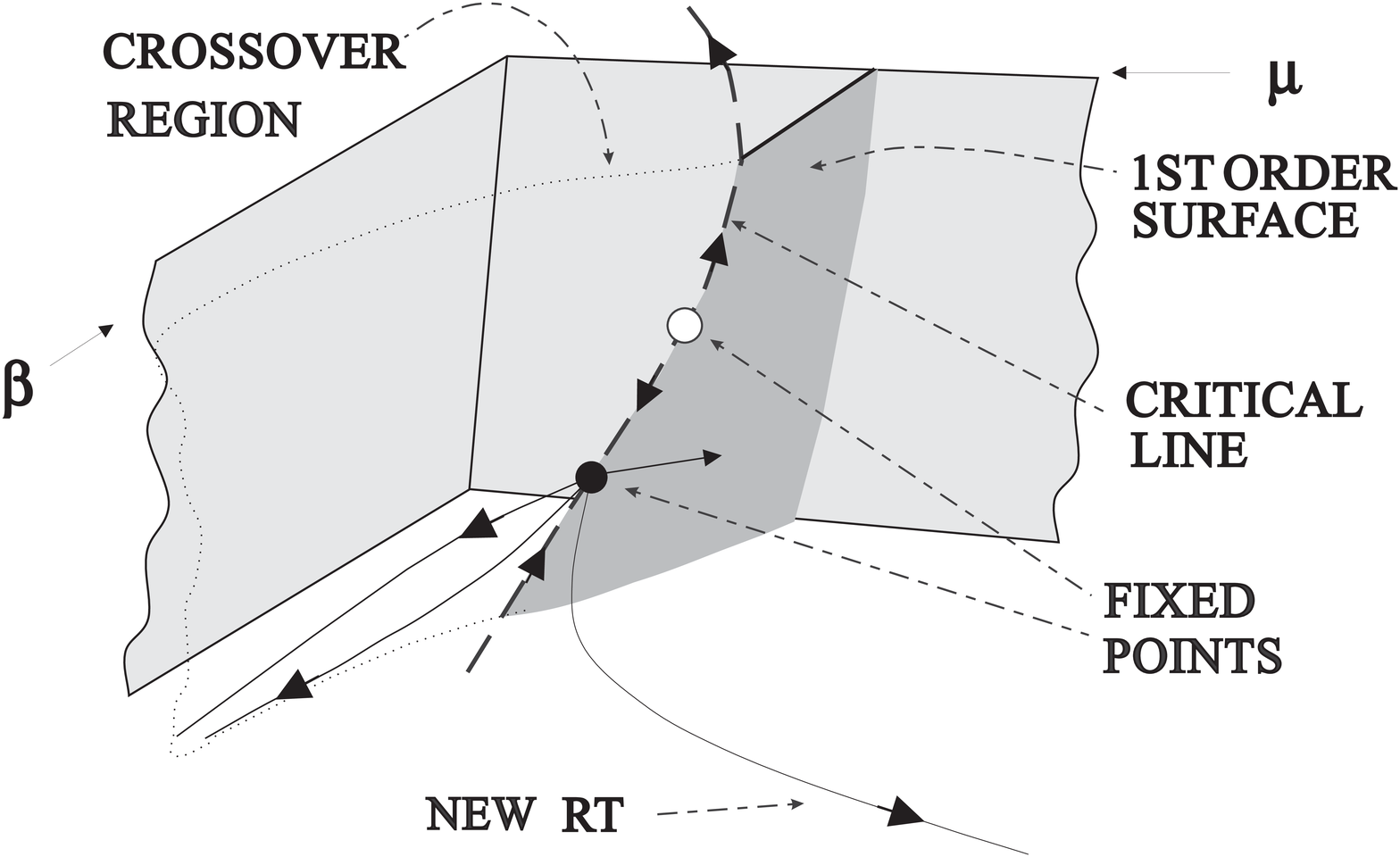,height=45mm}
\enc
\vskip -8 truemm
\caption{\label{f4} A line of critical points terminates the first-order surface and contains 
new fixed points from one of which flows the RT responsible for the observed cross-over
behaviour.}
\vskip -5truemm
\enf

\enlargethispage*{4pt}
In figure \ref{f4} we shown an artist's impression of a possible topology for the RG flows.

\end{document}